\documentclass[twocolumn,letterpaper,aps,preprintnumbers,longbibliography,notitlepage,longbibliography]{revtex4-1}
\usepackage[latin9]{inputenc}
\usepackage{amsmath}
\usepackage{amssymb}
\usepackage{graphicx}
\usepackage{esint}

\makeatletter

\pdfpageheight\paperheight
\pdfpagewidth\paperwidth

\pdfoutput=1

\usepackage{epsfig}
\usepackage{graphics}

\usepackage{mciteplus}
\mciteErrorOnUnknownfalse

\makeatother

\begin{document}

\preprint{FERMILAB-PUB-15-410-A}

\title{Exotic  Rotational  Correlations  in Quantum Geometry}

\author{Craig Hogan}

\affiliation{University of Chicago and Fermilab Center for Particle Astrophysics}

\begin{abstract}
It is argued that the classical local inertial frame used to define rotational states of quantum systems is only approximate, and that geometry itself must also be rotationally quantized at the Planck scale.  A Lorentz invariant statistical model of correlations in quantum geometry on larger scales predicts spacelike correlations that describe rotational fluctuations in the inertial frame. Fluctuations are estimated to significantly affect the gravity of quantum field states on a macroscopic scale, characterized by the Chandrasekhar radius.  It is suggested that the cosmological constant might be a signature of exotic rotational correlations entangled with the strong interaction vacuum, and have a value determined entirely by Planck scale quantum gravity and Standard Model fields.
 \end{abstract}
\maketitle

\section{introduction}

If, as is often supposed\cite{Rovelli2004,Thiemann:2007zz,Ashtekar:2012np,Mohaupt},  space-time emerges from a quantum system with finite information content,
it should display new  quantum correlations of geometry, that differ in character  from those of particles and fields.
Although the natural scale for its correlations is  the Planck length, $l_P\equiv c t_P\equiv \sqrt{\hbar G/ c^3}= 1.616\times 10^{-35}{\rm m}$, 
studies of gravitation\cite{Jacobson1995,Verlinde2011,Padmanabhan:2013nxa} suggest  that the  information content of any system is holographic, which would require nonlocal, spacelike correlations on all scales.  The character of these  ``exotic'' correlations is not known.

This paper addresses some physical consequences of a specific candidate form for these correlations.
It is based on a model for how quantum geometry approximates the classical local inertial frame that conventionally defines  absolute rotation, for example by measurements of centrifugal force or particle spin.
A precise Lorentz invariant statistical model\cite{Hogan:2016tkp}, based on Planck information density in proper time and spacelike Planck scale displacements that leave the light cones of any observer's  world line invariant, determines the specific form of exotic correlations in space and time, and their experimental signatures. Geometry becomes nearly classical on large scales, but  displays ``spooky'' nonlocal  quantum correlations in trajectories at spacelike separations that act like differential rotational fluctuations from the classical  inertial frame.

Although the precise model has only been formulated for  fixed causal structure, for  systems close to flat space it may serve as an adequate approximation for estimating new effects of geometrical degrees of freedom in holographic quantum gravity. It  is used here to show that nonlocal, spacelike  exotic rotational correlations  of geometry could   resolve well-known inconsistencies of effective field theory with gravity in the infrared.  In particular, it is argued that entanglement of geometry with the strong interaction vacuum could account for the observed  value of the cosmological constant.

\section{ Quantum indeterminacy of the inertial frame}

In classical relativity, the rate of change of any direction can be measured locally. 
As famously illustrated by Newton with a rotating bucket of water, a system with a constant orientation relative to absolute space experiences no centrifugal acceleration.  The local inertial frame  is defined 
for arbitrarily small systems, with no modifications to the concept of absolute space. 
 In the standard theory of  quantum matter, rotational states of  a  quantum system, such as a particle spin, are still defined relative to absolute, classical space-time down to infinitesimal scales. 
 
However, when the frame dragging (or Lense-Thirring) effect of general relativity is included, the gravity of any device used to measure rotation influences the inertial frame of the measured space. The standard separation of quantum matter  and classical space-time then becomes inconsistent. 
A quantum ``bucket'' is  in a superposition of rotational states with respect to classical space, but since each state has a  different flow of mass-energy, the  inertial frame is also placed into a quantum superposition.  
Quantum geometry  thus displays
exotic rotational correlations that  arise from  quantum degrees of freedom that are not included in any 
 standard  quantum theory.

To illustrate the effect, consider a  device to measure rotation on the Planck scale $l_P$. General relativity predicts that its {\it maximum} mass is that of a black hole of this size,  the Planck mass, $m_P= \sqrt{\hbar c/G}$, whereas quantum mechanics requires that its {\it minimum} mass is that of an elementary particle of this size, which is also the Planck mass. 
Thus, the device has about the  mass and size as a black hole, but it also has about the same mass and  size as a single elementary particle.   

According to quantum mechanics, its angular momentum has a value of $ \hbar$ times its (integer or half-integer) spin.  The spin is defined relative to the local inertial frame, but it does not have a definite value: its projection onto any axis is an operator,  characterized by a noncommutative spin algebra.  It can have a definite spin about at most one axis, determined by measurement; the other components are indeterminate  superpositions.

At the same time,  the  rotational energy flow and gravitational potential in a Planck scale device with angular momentum $\hbar$  have about the same values as those of a maximally spinning black hole.
According to general relativity,  a shell of mass $M$ and radius $r$ that rotates at an angular frequency  $\omega_M$ ``drags''  the  nearby inertial frame  at a rate 
$\omega\approx (GM/r)\omega_M$ compared to what it would be without the mass.
Thus, the gravity of the device causes  the local inertial frame to rotate by a substantial amount compared  to the distant universe---   comparable to the spin rate of a Planck mass black hole, with  frequency $\omega \approx t_P^{-1}$. 
But that means that the postulated set-up to measure spin in a local inertial frame is actually inconsistent: 
{\it The quantum indeterminacy of spin of the measuring device is inherited by the space itself}.  

%In other words, it is not possible to separate the quantum properties of the measurement device from the space whose local rotation it is supposed to measure. 

Thus, extrapolation from standard gravity and quantum mechanics  implies that  there is no locally determinable nonrotating frame  on the Planck scale; instead, measurements in rotation rate about any axis yield a variance  $\langle\Delta \omega^2\rangle^{1/2} \approx t_P^{-1}$ with respect to a classical space-time. 
The  absolute nonrotating inertial frame of classical  relativity does not exist at the smallest scales,  but can only be defined statistically, over a region much larger than the Planck length. 

This Planck scale quantization of rotation
 contrasts with   Wheeler's picture of  Planck scale space-time
 as ``quantum foam,''
an extrapolation of quantum fields and gravity to the  Planck scale that predicts    a roiling sea of virtual black holes with qualitative changes in causal structure and even topology. 
By contrast, in the model developed here,  an exact symmetry of the Planck scale system protects causal structure and prevents the creation of virtual black holes or gravitational potential fluctuations. Quantum geometrical degrees of freedom are required to have purely  rotational symmetry, so it is consistent to ignore curvature fluctuations at all scales. 

However, quantum geometry still produces exotic rotational fluctuations, even on scales much larger than the Planck length.
As discussed here, they should  produce new measurable physical effects,  even in space-times with vanishing curvature on large scales.
The measurement devices considered here to define the inertial frame do not resemble Newtonian buckets based on centrifugal acceleration. Instead, a covariant theory is developed based on light propagation.

\section{Exotic Rotational Correlations}  

\subsection{Statistical  Lorentz Invariance}

The concept of an observer-independent space-time is central to relativity, but is at odds with the quantum-mechanical principle that any measurement only has meaning in the context of an observer, that is,  the preparation of a state and its correlation with the state of a measurement apparatus.
Nevertheless, any theory of geometrical positions must preserve the principle of statistical Lorentz invariance---  that is, the  correlations predicted by a theory should  not depend on an arbitrary  choice of coordinates, or the frame used to describe them.  

%The non-observable quantities in the  model do not have to be invariant.  

This principle can be satisfied   by a statistical model\cite{Hogan:2016tkp} based on two principles:
(1) The system has a  Planck scale information density, in invariant proper time, on the world line of a measurement; and 
(2) The physical effect of the correlations can be represented by random Planck scale spacelike displacements that exactly preserve classical causal structure defined by the light cones of the world line of the measurement.
In this model, null intervals are protected, but  timelike and spacelike intervals  emerge as statistical approximations with exotic correlations and noise.
The detailed Planck scale quantum operators are not needed to compute the structure of exotic correlations.

This statistical description   is much more limited in scope than any theory of  quantum gravity. 
Indeed, the large scale correlations do not depend on any dynamics,  only on an invariant causal structure.
The correlations nevertheless lead to specific and distinctive measurable physical effects.
The model can be viewed as a step towards understanding how inertial frames with Lorentz symmetry emerge from a quantum system.

\subsection{Correlations on Light Cones}

\begin{figure}
\begin{centering}
\includegraphics[width=\linewidth]{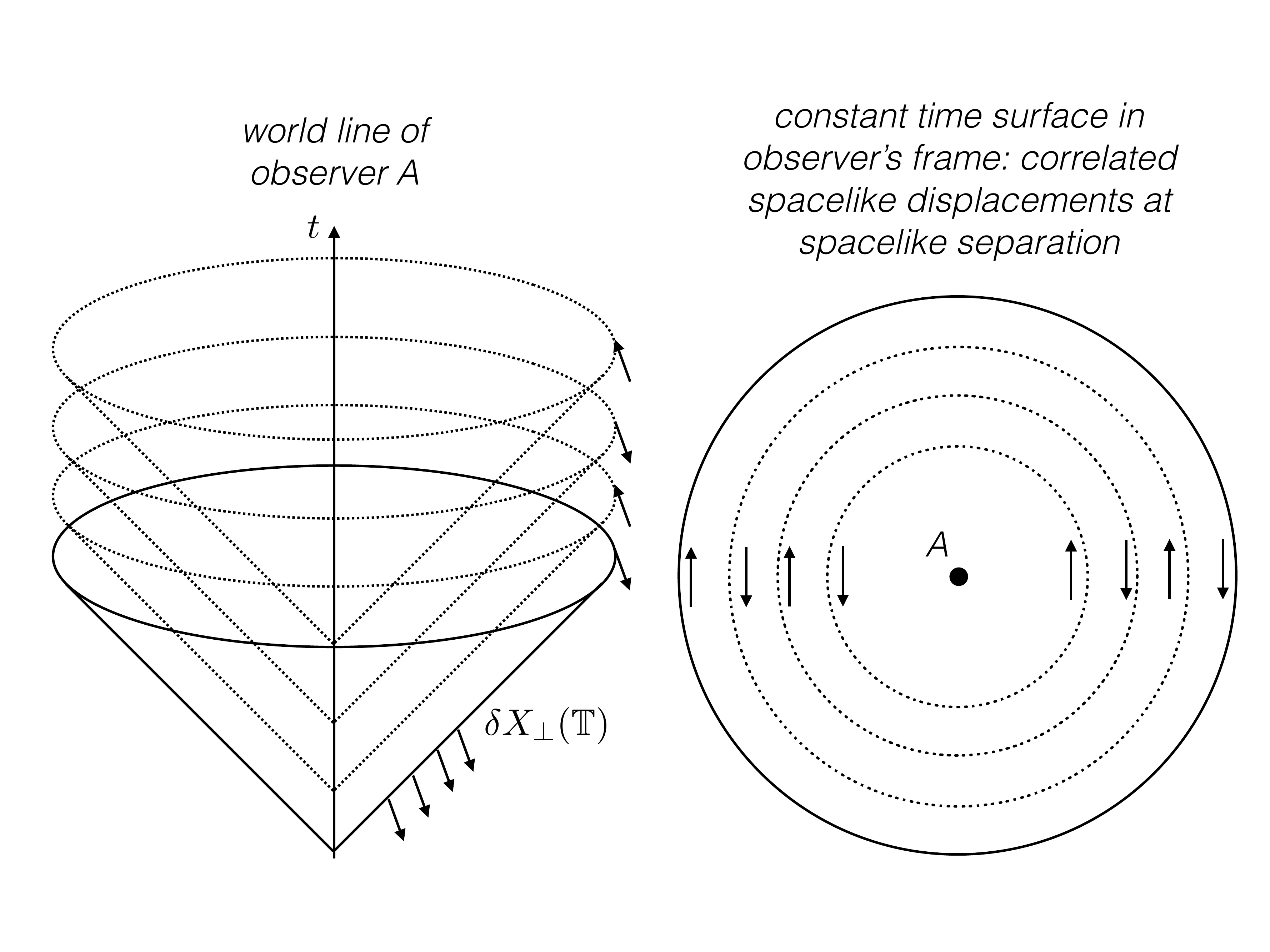}
\par\end{centering}
\protect\caption{Two views of exotic displacements in a twisted light-cone foliation of space-time.  Left, a 3+1D view, with one dimension suppressed, of light cones from a time series on an observer's world line.  Right, a 2D slice of the same system at a single proper time in the observer's frame shows a series of  nonlocally correlated differential displacements at spacelike separation. The information and causal structure in the twists agrees with holographic gravity if transverse Planck length displacements are highly correlated on light cones separated by less than a Planck proper time. These exotic rotational correlations extend indefinitely in all directions.\label{lightcones}}
\end{figure}

 A  statistically Lorentz invariant  model of large scale exotic correlations follows from the principle that causal structure is invariant for any observer.
The effects of Planck scale quantum geometry are modeled as a  transverse spatial displacement  represented as a random variable $\delta X_{\perp}$ of space-time position with  $\langle  \delta X_{\perp}^2\rangle = l_P^2$.   
Every event on the future light cone of a  Planck proper time interval on a world line $A$ is associated with  a  transverse displacement $\delta X_{\perp}$  with respect to  $A$.
The spatial covariance of the  displacements can be written as\cite{Hogan:2016tkp}:
\begin{equation}\label{eqn:covariance}
\langle \delta X_{\perp}(\mathbb{T}'_A) \delta X_{\perp}(\mathbb{T}''_A)\rangle_A=\begin{cases}
c^2t_{P}^{2}\;, & \left|\mathbb{T}'_A-\mathbb{T}''_A\right|<t_{P}\\
0\;, & \textrm{otherwise.}
\end{cases}
\end{equation}
Once an observer's world line has been specified, the proper time coordinate $\mathbb{T}_A$ is an invariant label for any event, so {\it the   correlations of $\delta X_{\perp}$  are entirely determined  by  the relative positions of events in an invariant causal structure.}\footnote{The artificially sharp boundaries of bins used to discretize the continuum in Eq. (\ref{eqn:covariance})  do not affect the form of the large scale correlation.  Variations of the discretization affect only the overall normalization, which is absorbed into the definition of the parameter $t_P$.} Thus,  the model written this way is manifestly Lorentz invariant on scales larger than $t_P$. 
The displacements can be visualized as   ``quantum twists of space''   on the 2-surfaces where light cones intersect  a constant-time hypersurface  (see Fig. \ref{lightcones}). 
As desired, a standard  classical nonrotating inertial frame emerges as the long time average of the displacements since by construction $\langle \delta X_{\perp}\rangle =0$. The correlations are both local, confined to a Planck proper time interval of light cone time, and spatially nonlocal, extending everywhere in space out to null infinity.

% If the speed of light is constrained to be locally constant, a physical measurement of  the inertial frame fluctuates from its classical value. 

\subsection{Physical Effects on Light and Clocks}

\begin{figure}
\begin{centering}
\includegraphics[width=0.8\linewidth]{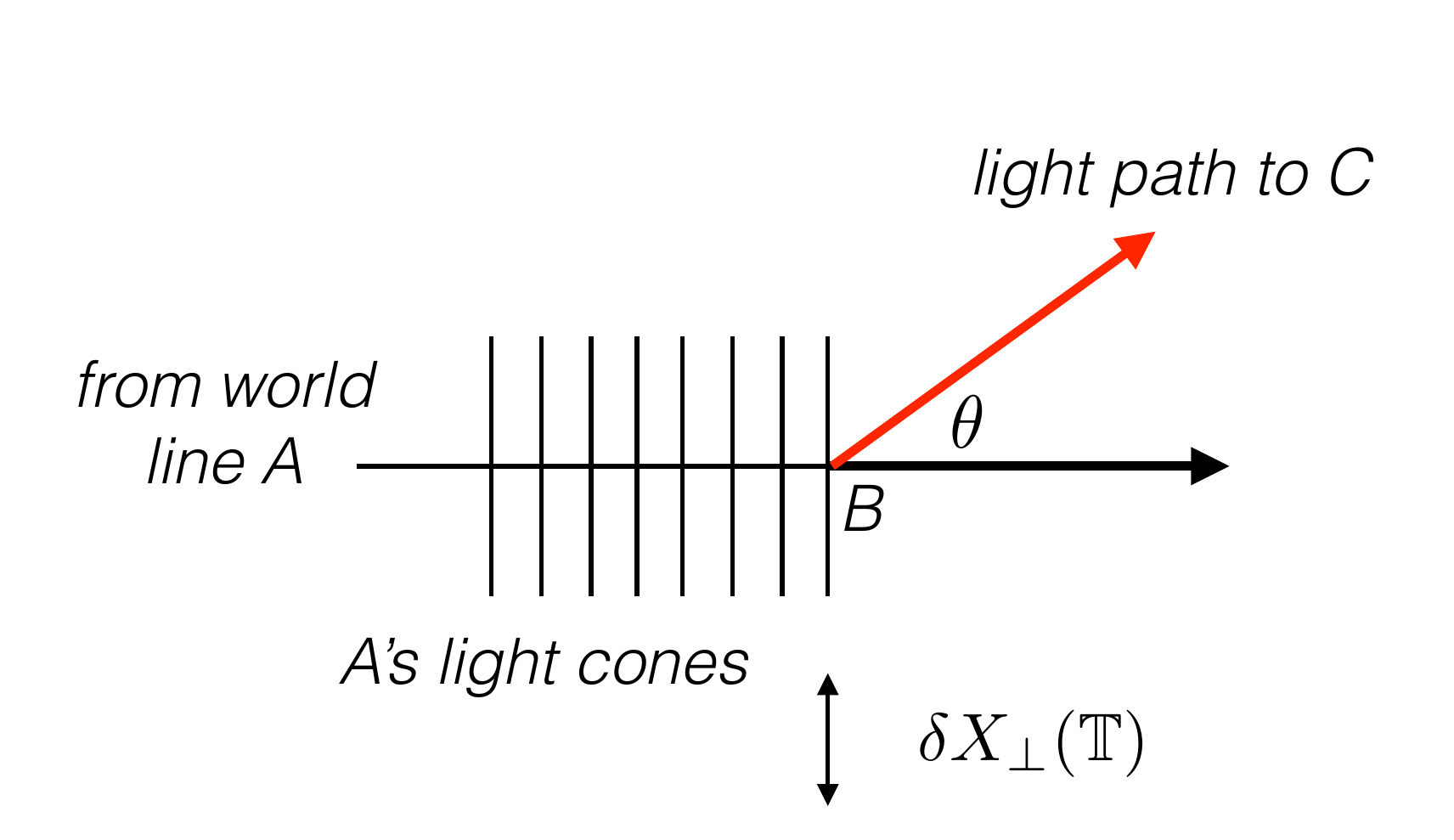}
\par\end{centering}
\protect\caption{ A light path near a world line $B$ is shown in one spatial plane at a single time, in the classical rest frame of a   distant inertial observer $A$ whose proper time $t_A$ represents laboratory time.  Light cones are represented by a series of null wave fronts with normal  at an angle $\theta$ from the light path direction. A  projection of Planck scale displacements leads to a  random drift (Eq. \ref{timevariance}) of light phase along the path in laboratory  time.
\label{projection}}
\end{figure}

The exotic  correlation leads to measurable physical effects.
Consider the exotic noise in the comparison of the rate of  arrival of Planck interval pulses from  observer $A$ with the phase of  light propagating along a path at some other point in space $B$, the world line of a distant body at rest in $A$'s classical inertial frame
(see Fig. \ref{projection}). Denote the angle between the light cone  surface normal and the light path in the  rest frame of $A$ by $\theta$. Assume that the local behavior of light, on a scale much larger than $l_P$ but smaller than the $AB$ separation, is not changed from standard physics.
Then an exotic random transverse position displacement $\delta X_\perp$   changes the location  of the  wavefronts of  light from their classical position, in their direction of propagation,  by a longitudinal displacement 
\begin{equation}\label{wavefront}
c \delta t = \delta X_\perp \sin(\theta).
\end{equation}
The effect on light depends only on the local angle $\theta$ between the light direction and the direction to $B$ from the observer. Equation (\ref{wavefront}) expresses the  physical effect of transverse  Planck scale displacements,  if the emergent local speed of light  is always $c$. 

The total  displacement  of light along the propagation direction $BC$ in $t_A$ time units is a sum of projected random transverse displacements,  $\Delta t_{BC}=\sum \delta t= \sum \delta X_\perp \sin(\theta)$.  The  variance  $S_0= \langle \Delta t_{BC}^2 \rangle$ between the observer's clock and that defined by displacement (or phase) along the the $BC$ light path accumulates at a rate in $A$'s proper time,
 \begin{equation}\label{timevariance}
d S_0 / dt_A = t_P^{-1} t_P^2  \sin^2(\theta) [1-\cos(\theta)].  
\end{equation}

The first factor  of $t_P^{-1}$ corresponds to the arrival rate of $A$ clock pulses at $B$. Each of the pulses represents a light cone from $A$, which  contributes  transverse displacement variance that appears   in the second factor of  $t_P^2$, from Eq. (\ref{eqn:covariance}). 
Each displacement  projects onto the light path according to standard geometry,  so the factor of $\sin(\theta)$ from Eq. (\ref{wavefront}) leads to a factor of  $\sin^2(\theta)$ in the variance.
The final factor of  $[1-\cos(\theta)]$  accounts  for  the  fact that the  wave fronts of  propagating light travel upstream or downstream relative to the clock pulses.
If we compare $A$ pulses with  light bouncing between $B$ and $C$ back and forth in both directions, the cosine contribution  averages to zero over many round trips.  However, there is still  an exotic drift between the two measures of time that depends only on the  orientation $\theta$. From the  point of view of a classical observer, this drift is attributed to exotic fluctuations in the inertial frame.

Notice that light reflected back to the observer has $\theta=\pi$, so there is no  exotic displacement of phase for radially propagating light, consistent with causal symmetry.   For other orientations, the clock drift rate only depends on $\theta$, so is independent of the $AB$ distance.
However,  its interpretation does depend on the adopted frame:
an observer at $B$ is not in an inertial frame relative to an observer at $A$, and {\it vice versa}.

%Note that $d S_0 / dt_A$  is invariant with respect to rotations around the $AB$ separation vector, so it is consistent with independent projections onto orthogonal planes.  

%Even ``freely falling''  clocks  appear to drift, in a way that depends on orientation.
%The fluctuation can be viewed as a fundamental  ``anisotropy of time''  that arises from its noncommutation with (transverse) spatial position at the Planck scale.  This strange behavior is needed to preserve exact causality as time and space emerge from the Planck scale. 

%Time is a discrete sequence along one world line (the observer's), but everywhere else, it cannot be separated from space.

\subsection{Experimental Observables}

The  variance in  Eq. (\ref{timevariance}) is well defined geometrically, but it is not a quantum  observable, because a  distant observer's clock is not locally measurable:  the quantity $S_0$ refers to a comparison of quantities at two different world lines, $A$ and $B$.
An observable correlation is a local comparison of arrival times or phases of light wave fronts  at a single world line.

For example, in the setup of Figure (\ref{projection}),  consider light that travels from $B$ to $C$ and back again along the same path.  At every point in the path the two directions have opposite signs for $\cos(\theta)$ in the integrand,  so the total round trip variance  $S_{RT}$, as measured at $B$ using $A$'s clock pulses,  is 
\begin{equation}
S_{RT}=  \int_{BCB} dt_A (dS_0/dt_A) = 2 t_P \int_{BC} dt_A \sin^2(\theta).
\end{equation}
In the case of  transverse propagation, $\sin(\theta)=1$, so the variance accumulates like a Planck random walk over a macroscopic distance. For a radial path, 
$\sin(\theta)=0$, there is no effect.

% In general, $\theta$ of course depends on $t_A$.

%\cite{Holo:PRL,Holo:Instrument}

% Interferometers are the only instruments currently capable of distinguishing exotic effects as small as those predicted. 

%Other techniques that measure relative spacelike positions averaged over many light crossing times have reduced sensitivity to the effects of fluctuating departures from a local inertial frame.

 A real world example is the signal of an interferometer, which measures light phase difference  between two paths that travel nonlocally through space, but begin and end at the same beamsplitter.  
A  statistical analysis of the exotic effect on propagating light in interferometers\cite{Hogan:2016tkp} makes exact predictions for signal correlation functions that depend only on the  shape of the light path.
 As seen above, the component of light that propagates in a transverse direction accumulates an exotic phase displacement  that grows  like a Planck random walk.   
The predicted amplitude of the effect is large enough to detect with the sensitivity already achieved by a correlated, superluminally sampled dual interferometer system\cite{Holo:PRL,Holo:Instrument}.

\subsection{Rotational Fluctuations}

Exotic rotational correlation can be visualized as   distance-dependent statistical fluctuations in rotation of the inertial frame.
The  fluctuations can be quantified by  the variance in longitudinal displacement of an entire  light path.   The accumulated variance over an infinite light path tangent to a sphere of radius $R$, from integration of  Eq. (\ref{timevariance}), yields $S_0= \pi t_P R/c$, dominated by the part of the path at separations not much larger than $R$. 
The directional  variation on scale $\approx R$ over a time $\approx R/c$ is then about
 \begin{equation}\label{Pdiff}
\langle\Delta\theta^2\rangle\approx S_0c^2/R^2 \approx l_P/R.
\end{equation}
Directions at separation $R$ fluctuate on timescale $R/c$ with a variance in rotation rate
\begin{equation}\label{OmegaR}
\langle\omega^2(R) \rangle \approx    c^2 l_PR^{-3}.
\end{equation}

The exotic directional fluctuations and  information content  in this  model qualitatively agree with earlier extrapolations of gravity and quantum mechanics\cite{Hogan:2010zs,Kwon:2014yea,Hogan:2015kva}  based on causally constrained wave solutions with  Planck frequency bandwidth, or other holographic bounds on information.
Here,  the added constraints of statistical Lorentz invariance and exact causal symmetry remove  ambiguities in predictions for the spatial character of the states and fix the relationship between emergent space and time.  The only  parameter is the overall normalization fixed by the physical value of the Planck length, which is  given by gravitational theory.

\section{Gravitational effects}
Exotic rotational displacements in this model are based on a fixed, classical causal structure of  infinite flat space-time. 
The model should precisely predict correlations in interferometers, which are much smaller than the gravitational radius of curvature.

Because the model does not include any matter, dynamics or curvature, it  does not describe quantum gravity. However, 
in a thermodynamic interpretation of gravity\cite{Jacobson1995,Verlinde2011,Padmanabhan:2013nxa},
the  quantum degrees of freedom of flat space-time are the same as  those of curved space-time; they represent different configurations of the same geometrical quantum system.
The  number of degrees of freedom in our model associated with a proper time interval, the area bounding its invariant causal diamond in Planck units, has been normalized to agree with the holographic information content of gravity.
In this picture, the flat-space model describes the degrees of freedom of quantum gravity for systems close to the ground state.
It can  be extrapolated to estimate how rotational correlations affect quantum states of fields, and how field excitations affect  emergent macroscopic causal structure.
The extrapolations suggests specific mechanisms to resolve  two related infrared gravitational catastrophes of virtual field states: how the vacuum avoids fluctuating spontaneously into black holes \cite{CohenKaplanNelson1999},  and how its effective mean gravitating density comes to have a nonzero value vastly smaller than the characteristic energy density of virtual field states\cite{Weinberg:1988cp}.

\subsection{Gravity of Virtual Field States}

Exotic rotational correlations in quantum field states  add  new correlations in the infrared that  can resolve long standing  infrared inconsistencies of effective field theory with gravity in  large volumes\cite{CohenKaplanNelson1999}.
A  quantum field system includes all possible states of a field, including nonzero occupation numbers for all of its modes. A free field up to some ultraviolet cutoff  scale with wavenumber $k= m c/\hbar$ has  about $ (Rk)^3$ independent modes in a volume of size $R$. In a state where each  mode has mean occupation number of order unity, the number of particles per volume is about $( mc/\hbar)^3$.
The  energy of the particles in this state matches the gravitational binding energy in a macroscopic volume  at an idealized    Chandrasekhar radius\cite{1931ApJ....74...81C}:
\begin{equation}\label{CradiusP}
 R_{C}/l_P \approx (m_P/ m)^2,
\end{equation}
where   $m_P=\sqrt{\hbar c/G}$ denotes the Planck mass.
In a volume with a size larger than $R_{C}(m)$, the virtual field state has a mass larger than that of  a black hole of the same size, which is of course an impossible physical state,  inconsistent with general relativity.  

In ref. \cite{CohenKaplanNelson1999}  it was suggested  that new kinds of  quantum correlations somehow  prevent this infrared catastrophe.
Exotic correlations in the background geometry provide a specific mechanism to accomplish this: the correlated twists of light cones add spacelike correlations to field amplitudes  on large scales that would be  independent in a standard classical background.
Using Eqs. (\ref{Pdiff}) and  (\ref{CradiusP}),  the accumulated twist matches a particle wavelength for  volumes larger than  $R_C(m)$, and thereby naturally leads to an effective cut-off at the right scale. In the field-state description, it prevents the catastrophe by reducing the effective number of degrees of freedom. The overall state of the combined system (fields and geometry) can then in principle have a consistent quantum description.

\subsection{Cosmological Constant}
  The observed acceleration of the cosmic  expansion\cite{Riess:1998cb,Perlmutter:1998np,Frieman:2008sn}
 can be interpreted in  general relativity as an effect of a nonzero cosmological constant
   $\Lambda$ in  Einstein's field equations, whose value is  precisely measured from cosmological data.
  A nonzero value of  $\Lambda$ produces a positive acceleration, in the absence of other sources of gravity, proportional to the separation $r$ of two bodies in Newtonian coordinates,
   \begin{equation} \label{lambda}
\ddot r = H_\Lambda^2 r,
\end{equation}
where  $ H_\Lambda^2\equiv \Lambda/3$.  The same effect can be interpreted  as a ``dark energy'' of a field vacuum,  with a gravitating density $\rho_\Lambda = 3 H_\Lambda^2/8\pi G$. If interpreted in terms of a standard effective potential, the  field responsible for the observed cosmic dark energy must have a scale  wildly different from those of known fields. Moreover, a straightforward estimate of the   energy density  of  vacuum fluctuations in field amplitude with a  Planck scale cutoff  yields a dark energy density or value of $\Lambda$ that is too large by a factor of about $(H_\Lambda t_P)^{-2}$---   about 122 orders of magnitude\cite{Weinberg:1988cp}. 

There are thus at least two problems to solve: why $\Lambda$ is so small in Planck units,  and why it does not exactly vanish.
The exotic rotational correlations may solve both of these problems: the causal symmetry of quantum gravitational degrees of freedom could account for the near vanishing of gravity from field fluctuations, while at the same time,  the  vacuum states of known fields  could slightly break the scale invariance of  quantum gravity in the right way, and  by  the right amount,  to  account for the observed cosmic acceleration.

In the thermodynamic interpretation  of general relativity, curvature of any kind is a collective phenomenon  that only acquires meaning on scales much larger than the Planck length, and gravity occurs as a statistical behavior in an excited system\cite{Jacobson1995,Verlinde2011,Padmanabhan:2013nxa}.   The ground state is a flat space-time, with the possible addition of an arbitrary cosmological constant. 
This idea is consistent with our flat-space model of exotic rotational correlations:    quantum fluctuations of geometry  have no effect on causal structure, and therefore produce no curvature. 
If  the same symmetry applies to the degrees of freedom of quantum gravity, 
 excitations associated with virtual  particles would still preserve local causal structure, so that the ground state curvature of the vacuum nearly vanishes.
The causality-preserving symmetry of purely rotational degrees of freedom  would  naturally account for a near-absence of  gravity from vacuum  field fluctuations, in the same way that it suppresses foamlike fluctuations  of curvature and topology at the Planck scale.

At the same time,  small effects of a  field vacuum would be expected in quantum gravity that are not included in the flat model. As seen above, because correlated light cone twists extend indefinitely in space-time,  exotic rotational correlations significantly alter trajectories on macroscopic scales $\approx R_C$, where the Planck holographic information of space-time becomes comparable to that in fields. In the cosmic context,  rotational degrees of freedom, when entangled with  field vacuum degrees of freedom, could  lead to cosmic acceleration.   

% large scales  information content  global causal structure  future event horizon.

Ordinary classical rotation creates a  kinematic centrifugal acceleration  that in some ways resembles cosmic acceleration.
In  a classical  system rotating at a rate $\omega$, a body at separation $r$ from the axis of rotation experiences a centrifugal acceleration
\begin{equation}\label{centrifuge}
\ddot r = \omega^2 r.
\end{equation}
Like cosmic acceleration, it is  proportional to $r$, and always positive. 
It also affects all bodies equally, independent of mass or other properties: it depends only on position.

In the case of exotic rotational fluctuations,  the time and space averages $\langle \omega\rangle$ vanish, so the anisotropy  of acceleration associated with  rotation around a particular axis, and the inhomogeneity associated with random spatial variations in $\omega$,  average to zero in a large system. 
However, exotic rotational fluctuations still have $\langle\omega^2\rangle>0$, so  if they represented real classical  motion they would produce a spatially and temporally fluctuating centrifugal acceleration. The time averaged radial component $\langle \ddot r/r\rangle $ depends on scale; from Eqs.  (\ref{lambda}), and (\ref{centrifuge}),   it equals cosmic acceleration  on the scale $R_\Lambda$ where variance (Eq. \ref{OmegaR}) is 
\begin{equation}\label{Rlambda}
\langle\omega(R_\Lambda)^2\rangle =  H_\Lambda^2=\Lambda/3.
\end{equation}
 The integrated radial component of  exotic rotational fluctuations thus statistically mimics cosmic acceleration on large scales.
The observed cosmic acceleration would naturally arise if  centrifugal acceleration in the emergent space-time  remains ``virtual'' at separation scale $R< R_\Lambda$, but  behaves like a real fluctuation with $\langle\ddot r/r \rangle\approx \langle\omega(R)^2\rangle $ at separation scale $R>R_\Lambda$.  Heuristically, we could say that the rotationally fluctuating vacuum on the scale $\approx R_\Lambda$ statistically  ``shakes space apart'' on large scales.

Notice now a possibly profound coincidence,  that the actual value  of   $R_\Lambda$ needed to explain cosmic acceleration emerges naturally from the Standard Model vacuum. From Eqs. (\ref{OmegaR}) and (\ref{Rlambda}), we find
\begin{equation}\label{RLambda}
R_\Lambda / l_P \approx   (H_\Lambda t_P)^{-2/3},
\end{equation}
the Chandrasekhar  radius (Eq. \ref{CradiusP}) for   particle mass 
\begin{equation}\label{coincidence}
m_\Lambda/m_P\approx (R_\Lambda/l_P)^{-1/2}\approx  (H_\Lambda t_P)^{1/3}.
\end{equation}
 Remarkably, this particle mass scale $m_\Lambda$ derived  from cosmic acceleration is about equal to the QCD chiral symmetry breaking scale $m_Q\approx 200 {\rm MeV}/c^2$,  where states of the strong interaction vacuum change from massless to massive behavior.
 Cosmic acceleration with the observed properties occurs if,  in quantum gravity,  exotic rotational fluctuations of  vacuum states with wavenumber $\le m_Q c/\hbar$ create real centrifugal fluctuations of order $\langle \omega(R_C)^2\rangle$ at their Chandrasekhar radius, $ R_C(m_Q)$. 
 A cosmic horizon forms on the scale $c/H_\Lambda$ from the integrated effect of many small,  $R_\Lambda$-scale regions fluctuating at   angular velocity $\omega\approx H_\Lambda$. 
 Although the cosmological constant is predicted to fluctuate in this model, the  scale  $R_\Lambda\approx R_C(m_Q)\approx 60$ km  is so small  that the predicted spatial and temporal fluctuations  make no detectable difference from a uniform, classical cosmological constant. 

%It is natural to identify this also as the scale where  the Lorentz invariance  of  the field vacuum is broken by the comoving cosmic frame, defined by world lines of  observers who see  statistical isotropy. 

The coincidence of  kinematic behavior and scale hints at a    physical connection   between the cosmological constant and  strong interactions, as contemplated long ago by Zeldovich\cite{Z68}, and  more recently by Bjorken and others\cite{Bjorken:2010qx,Bjorken:2002sr,Randono:2008wt,Carneiro:2003zw,Schutzhold:2002pr}.   
Causal symmetry provides a physical rationale for why the QCD vacuum might make kinematic effects of exotic rotational fluctuations real  at  the scale where virtual particles acquire mass, but remain virtual at scales where they are massless.
Classical centrifugal acceleration from rotation at rate $\omega$ relative to the inertial frame can be interpreted in the rotating frame as
a linear radial gradient of time dilation,  an apparent curvature with radius $c/\omega$.
The exotic  shift in  vacuum phase around a closed loop of dimension $\approx R_\Lambda$ creates a  displacement  $\approx \hbar/m_{Q}c^2$. This shift emerges  as  a real gravitational time dilation between  world lines with this separation, because the QCD vacuum at low energies, unlike the light paths  considered above, does not obey the null symmetry of the geometry.
Massive  particles can form  clocks and local oscillators, which can measure exotic radial time dilation between separate clocks in the rotating frame.  That cannot happen with particles moving only on null trajectories;  photons on their own cannot  be used to measure time.

Although the full quantum gravity system presumably obeys unitarity,  the QCD field subsystem on its own does not appear to conserve information, because it is entangled with geometry via exotic departures from classical time, such as those discussed above.  Timelike  states  at low energy cause information to be lost from  fields  on the  scale $R_\Lambda$,  where it is ``swallowed'' by the geometry.

The measured value of $\Lambda$  can thus be compared directly with  measurements of  $m_Q$ in terms of information flow.  The gravitational holographic entropy, or information ``lost'' over the cosmic horizon, should match a tiny departure from unitarity defined in classical time at the level of field states. 
  Suppose  that  some wavenumber $k_Q$  marks a sharp boundary between timelike and null information in  field degrees of freedom.
 The  holographic entropy $S$ is  one quarter of the area of the cosmic event horizon in Planck units:
 \begin{equation}
S= \pi t_P^{-2} H_\Lambda^{-2}.
\end{equation}
Dividing by the 3-volume gives the holographic information density,
${\cal I}_\Lambda = S (3H_\Lambda^3 /4\pi c^3) = 3 H_\Lambda/4 t_P^2c^3$.
The  density of free field  modes per 3D volume with an ultraviolet cutoff at wavenumber $k$ is
${\cal I}_f(k)=k^3 4\pi/ 3(2\pi)^3$.
A free scalar field   therefore matches cosmic information (that is, ${\cal I}_\Lambda= {\cal I}_f(k)$) for a field cutoff at
$k_Q= k_\Lambda \equiv ( H_\Lambda 9\pi^2/2)^{1/3}$.
Taking an estimate\footnote{This estimate of $H_\Lambda=  \Omega_\Lambda^{1/2} H_0$, where $H_0$ denotes the current value of the Hubble constant and $\Omega_\Lambda$ denotes the density parameter of the cosmological constant, corresponds to $h_0= 0.73\pm 0.017$ and $\Omega_\Lambda= 0.7\pm 0.01$ in standard cosmological notation.} of $H_\Lambda=  \Omega_\Lambda^{1/2} H_0= 0.99\pm 0.018 \times 10^{-61} t_P^{-1}$ from a typical fit to current cosmological data\cite{Riess:2016jrr,Olive2014}, we find that the cosmic information matches  field information with a cutoff at
\begin{equation}
k_Q c \hbar=  1.65\pm 0.01 \ \times 10^{-20} m_Pc^2= 201\pm 1.2\   {\rm MeV},
\end{equation}
remarkably  close  to $m_Q$  estimated from laboratory  measurements\cite{Olive2014}.

These physical arguments are based on comparing separate extrapolations from quantum field theory in classical space-time,  and exotic rotational correlations in flat, empty space-time.
They hint that  a full quantum theory that includes both fields and space-time might provide a  link between the QCD vacuum in its field-theory limit, and the cosmological constant in its general-relativity limit.
It appears that if exotic rotational correlations are shown experimentally to exist, an explanation based on conserved information in the whole system could naturally account for the absolute value of the cosmological constant from already known constants of physics,   without  additional parameters, scales, or fields.

Because $m_Q$ also  approximately determines atomic masses, the  cosmic acceleration timescale $H_\Lambda^{-1}$ in this scenario depends on approximately the same combination of physical constants  as a stellar lifetime\cite{Hogan:1999wh}--- albeit, for very different physical reasons.  
Thus, there may be a natural solution to what is sometimes called the ``why now'' problem of dark energy, that does not need to invoke anthropic selection from a multiverse.  In this picture, all  of the cosmic large numbers can be attributed to  logarithmic running of coupling  with energy scale\cite{Wilczek1999}.

 \begin{acknowledgments}
This work was supported by the Department
of Energy at Fermilab under Contract No. DE-AC02-07CH11359. 
The author is grateful  for many helpful comments from the Fermilab Holometer team, and for the hospitality of the Aspen Center for Physics, which is supported by  National Science Foundation Grant No. PHY-1066293.  
\end{acknowledgments}

  \bibliographystyle{apsrev4-1.bst}
\bibliography{bib_exotic}
 
\end{document}